\documentstyle[epsfig,psfig,emulateapj]{article}

\lefthead{RAO ET AL.}
\righthead{RAPID STATE TRANSITIONS IN GRS~1915+105}

\slugcomment{Accepted for publication in  The Astrophysical Journal}

\begin{document}

\title{RAPID STATE TRANSITIONS IN THE
GALACTIC BLACK HOLE CANDIDATE SOURCE GRS~1915+105 }

\author{ 
	 A.~R.~Rao\altaffilmark{1},
	J.~S.~Yadav\altaffilmark{1}, and 
 	 B.~Paul\altaffilmark{1,2}}
\altaffiltext{1}
{Tata Institute of Fundamental Research,
Homi Bhabha Road, Mumbai 400 005, India}
\altaffiltext{2}
{Institute of Space and Astronautical Science
3-1-1 Yoshinodai, Sagamihara, Kanagawa 229-8510, Japan\newline
arrao@tifr.res.in, jsyadav@tifr.res.in, bpaul@tifr.res.in}

\begin{abstract}
We examine the X-ray spectral and temporal behavior of
the Galactic black hole candidate source GRS~1915+105 
during its spectral state transition observed in 1997 March - 
August. The source was making a slow transition
from a low-hard state to a high-soft state in about three months and
during this transition it was exhibiting a series of fast variations
which can be classified as bursts. During one type of bursts, called the
irregular bursts, it was found that the source makes rapid transitions
between two intensity states. 
We have analyzed
the RXTE PCA data obtained on 1997 June 18 (when the source was making
rapid state transitions) and compared the results with two sets of data
each pertaining to the low hard-state and high-soft states.
We find that the power density spectrum (PDS) during the burst shows
remarkable similarity to that seen during the high-soft state of the 
source. The PDS during the quiescence, on the
other hand, is quite flat and is very similar to the PDS commonly seen during
the canonical low-hard state of black hole X-ray binaries. 
A low frequency QPO at $\sim$3 Hz is observed during the
quiescence, a property seen during the low-hard
state of the source.  The X-ray spectrum of the source 
during quiescence is similar to the
spectrum seen during the low-hard state of the source and 
the spectra during the burst and high-soft state are
very similar to each other.
These observations corroborate the suggestion made by
Yadav et al. (1999) that during the
irregular bursts the source shows spectral state variations
at very fast time scales. 
We explain such fast state changes using the two component
accretion flow model in which the accretion disk
consists of a geometrically thin (and optically thick) standard disk
co-existing with an advection dominated hot thick disk.
\end{abstract}
\keywords{accretion, accretion disks --- binaries: close ---
black hole physics --- stars: individual (GRS~1915+105)  --- 
X-rays: bursts --- X-rays:  stars}

\section{INTRODUCTION}

The X-ray transient GRS~1915+105 was discovered with the WATCH all sky
X-ray monitor (Castro-Tirado et al. 1992) and it has been exhibiting 
a wide variety of temporal variability in its X-ray emission.
It has been identified with a superluminal radio source
(Mirabel \& Rodriguez 1994) and 
attempts have been made to connect the jet production with the
hard X-ray emission from the accretion disk (Harmon et al. 1997).
A simultaneous X-ray and infrared observation of the source established
a close link between the non-thermal infrared emission (presumably coming
from a ``baby-jet'' analog of the superluminal ejection events) and the 
X-ray emission from the accretion disk
(Eikenberry et al. 1998). A diverse variability in the X-ray
emission, episodes of superluminal jet emission and evidences of disk-jet
symbiosis has made GRS~1915+105 an ideal source to examine several
concepts of accretion physics, black-hole environment, disk-jet connection
etc.

Greiner et al. (1996) presented the X-ray   variability characteristics
of GRS~1915+105 using the ROSSI XTE data which included repeated patterns
of brightness sputters, large amplitude oscillations, fast oscillations
and several incidences of prolonged lulls. Narrow quasi-periodic oscillations 
(QPO)
in the X-ray emission were discovered from the source using the Indian 
X-ray Astronomy Experiment (Agrawal et al. 1996) and the RXTE (Morgan 
\& Remillard 1996). Chen, Swank \& Taam (1997) found that the narrow QPO 
emission
is a   characteristic feature of the hard branch and it is absent in the
soft branch which corresponds to the very high state similar to those of
other black hole candidates. Morgan, Remillard,  \& Greiner (1997) 
made a systematic study of the different QPOs seen in GRS~1915+105
which ranged from 3 mHz to 67 Hz and based on the power spectra, light
curves and energy spectra 
classified the X-ray emission into four different emission states, none of
which appear identical to any of the canonical states of black hole 
binaries. Paul et al.  (1998b), however,  found that the 0.5 -- 10 Hz QPO traces
the change of state from a ``flaring state'' to a low-hard state quite
smoothly along with other X-ray characteristics like the low frequency
variability. Trudolyubov, Churazov, \& Gilfanov (1999) studied the  1996/1997
low luminosity state and state transitions using the RXTE data and
concluded that the QPO centroid frequency is correlated with the spectral
and timing parameters and these properties are similar to other
Galactic black hole candidates in the intermediate state. Muno, Morgan,
\& Remillard (1999) sampled the RXTE data over a wide range of properties
and found that the 0.5 -- 10 Hz QPO can be used as a tracer of the spectral
state of the source and the source mainly stays in two states: the spectrally
hard state when the QPO is present and the soft state when the QPO is absent.

The X-ray intensity variations seen in GRS~1915+105 were classified 
 as lull, flare and
splutter in Greiner et al. (1996), and each of these states last for 
a few hundred seconds. Similar X-ray light curves were analyzed in detail by 
Belloni et al. (1997a; 1997b) who classified them as ``outbursts''; they found that
the source shows distinct and different spectral and temporal characteristics
during the ``quiescence'' and outburst. Taam, Chen, and Swank (1997)
detected a wide range of transient activity including regular bursts with
a recurrence time of about one minute and irregular bursts.  Belloni
et al. (1997b) made a detailed spectral analysis during a sequence of
``bursts'' and discovered a strong correlation between the quiescent
phase and burst duration. Paul et al. (1998a) detected several 
types of bursts using the IXAE data and found evidence for matter
disappearing into the event horizon of the black hole. Yadav et al. (1999)
made a systematic analysis of these bursts and classified them based
on the recurrence time. 

 Yadav et al. (1999) 
presented a  comprehensive picture for the origin  of these  bursts
in the light of the recent theories
of advective accretion disk.  It was suggested that the peculiar bursts 
are characteristic of the change of state of the source. 
The source can switch back and forth between the low-hard state and
the high-soft state  near  critical accretion rates in a very short time 
scale, giving rise to the irregular and quasi-regular  bursts.  The fast 
time scale for the transition of the state is explained by invoking the 
appearance and disappearance of the advective disk in its viscous time 
scale.  The periodicity of the regular bursts is explained by matching  
the viscous time scale with the cooling time scale of region beyond 
a shock front or a centrifugal barrier (Chakrabarti \& Titarchuk 1995).

In this paper we present the results of a study of the
time variability and spectral characteristics of the source GRS 1915+105 during  the
irregular bursts. We show that during these bursts the source makes distinct
transitions between the two spectral states in a few seconds.
In section 2 we present results obtained
from an  analysis of the data obtained from the Proportional Counter
Array (PCA) of the Rossi X-ray Timing Explorer (RXTE).
 In section 3 we discuss the importance of our results and in the last
section a summary of the results is given.

\begin{deluxetable}{lllclc}
\footnotesize
\tablecaption{The details of selected RXTE observations on  
GRS~1915+105 \label{tbl-1}}
\tablewidth{0pt}
\tablehead{
\colhead{PID} & \colhead{Date} &   \colhead{Start time} & \colhead{Duration (s)} & \colhead{Comments}  & \colhead{References} }
\startdata
20402-01-21-00 & 1997 Mar 26  & 20:03:27 &   3312 & Low-hard state & 1,2  \nl
20402-01-27-01 & 1997 May  8  & 16:12:28 &   1808 & Low-hard state & 3  \nl
20402-01-33-00 & 1997 Jun 18  & 12:58:13 &   $~~$3528  & Irregular bursts & 3,4 \nl
20402-01-38-00 & 1997 Jul 20  & 10:13:10 &   7520 & High-soft state  & 1 \nl
20402-01-41-00 & 1997 Aug 19  & 05:44:09 &   2592 &  High-soft state  & 1,3 \nl
\enddata
\tablenotetext{}{References: 1.  Muno et al. (1999); 2. Trudolyubov et al. (1999); 3. Yadav et al. (1999); 4.  Belloni et al. (1997) }
\end{deluxetable}

\section{DATA SELECTION AND ANALYSIS }

GRS~1915+105 was in a low-hard state during 1996 December to 1997 March 
when the hard X-ray spectral index ($\sim$2.0) and 
the soft X-ray flux (300 - 500 mCrab) were low (\cite{grei:98}; 
\cite{trud:98}). The source started a 
new outburst around 1997 April-May when the soft X-ray flux started 
increasing and the X-ray spectrum  became soft (spectral index increased 
to 3 $-$ 4). It reached the high-soft state in August 1997.
The 1.3 to 12.2 keV X-ray light curve of the source obtained from the
RXTE ASM archives is shown in Figure 1, from 1997 January to 1997 September.
  Individual dwell data
(see Levine et al. 1996) are plotted against the ASM day numbers
(which is equal to MJD - 49353).  A few dates (in 1997) are
marked on the top of the figure.
There are about 20 dwells
on the source per day, lasting for about 90 s. Data are in ASM counts
s$^{-1}$ (1 Crab = 75 ASM counts s$^{-1}$). 

\placetable{tbl-1}

The RXTE public archive contains several observations on GRS~1915+105 using the
RXTE PCA (Jahoda et al. 1996). These observations typically last for a few
thousand seconds and the start times of these observations are marked as
circles in Figure 1. Several of these observations are analyzed and reported
in the literature and these are marked with vertical arrows in the figure.
The observation times of the regular bursts reported as `rings' in the PCA 
color-color diagrams by Vilhu \& Nevalainen (1998) are marked with a 
`v' in the figure and the `irregular bursts' reported by Belloni et al.
(1997b) is marked as `B'. The peculiar repeated observation of the 
`kink' in the light curve followed by a `lull' and then ringing flares
reported by Eikenberry et al. (1998) is marked by an `E'. Similar episodes of
flares have seen by Markwardt et al. (1999) and it is marked with an `M'.
The remaining observations which are reported in the literature are
from Trudolyubov, Churazov, \& Gilfanov (1999)  and Muno et al. (1999). 
 The observation times of GRS~1915+105 by the PPCs on board IXAE (Paul
el al. 1998a; Yadav et al.  1999) are marked as stars in Figure 1.  

\begin{figure}
\centering
\psfig{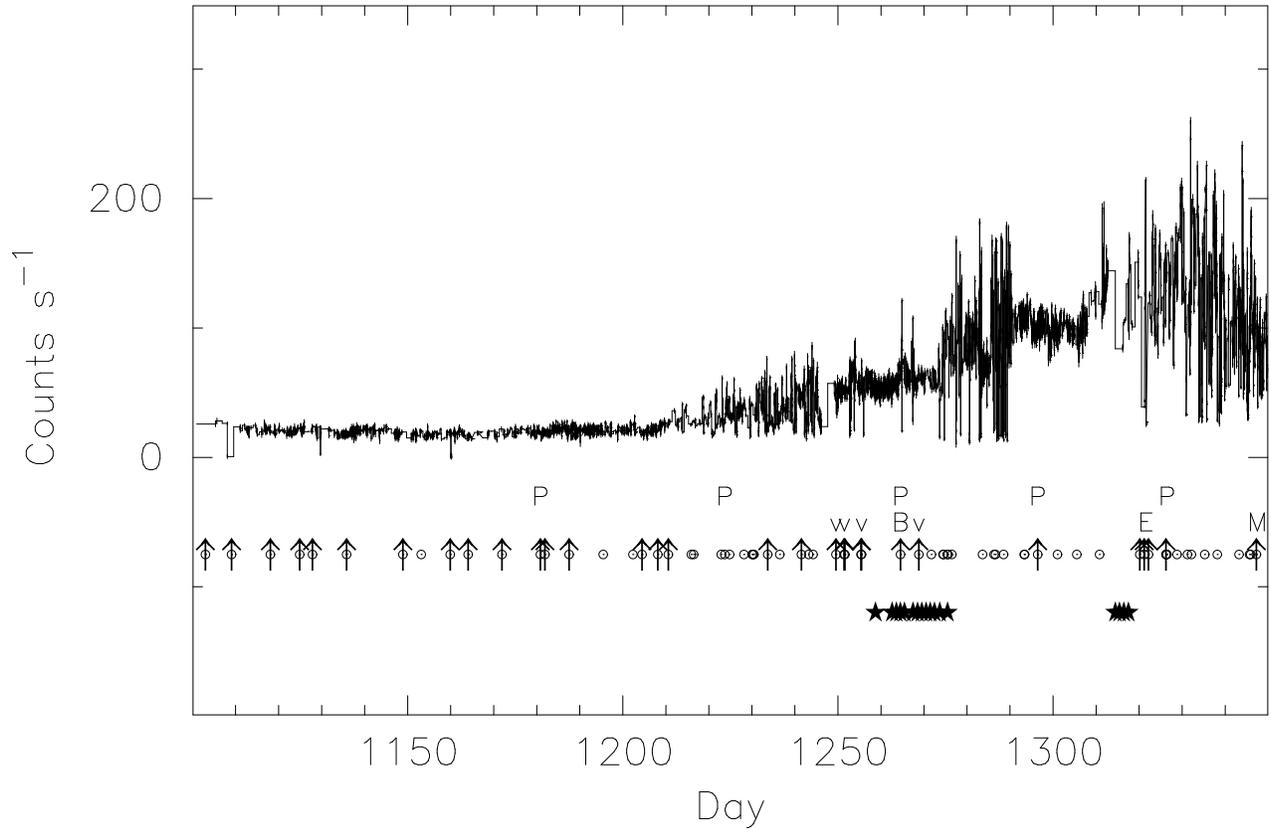}
\caption[fig1.eps]{The 1.3 $-$ 12.2 keV X-ray light curve of  GRS~1915+105 
obtained with the RXTE-ASM during 1997 January to  September when the source
made a slow transition from a low-hard state to a high-soft state. 
The X-axis is in ASM day numbers
(MJD - 49353.0006966).  A few dates (in 1997) are shown on the top of the
figure.
The RXTE pointed observations are marked as circles at the bottom of the
figure and those of which are analyzed and reported in the literature are
marked by vertical arrows, some of which indicated with alphabets (see text).
The RXTE pointed observations selected for the present study are marked with 
the letter `P'. The times of IXAE observations (Paul et al. 1998a; 
Yadav et al. 1999) are indicated by stars.}\label{fig1}
\end{figure}

\placefigure{fig1}

It can be seen from the figure  that the source was in a stable low-hard state
up to 1997 April 25 (day number 1210) when the average ASM count was
20 s$^{-1}$. Barring three episodes of dipping behaviors (which could be 
absorption dips seen in other black hole candidate sources like
4U 1630-47 - see Kuulkers et al. 1998), the source flux
was stable with  a rms deviation of 15\%. 
It should be noted here that the ASM data, with a dwell time of 90 s
and about 20 dwells per day, is insensitive to short term variabilities.
The spectral and temporal behavior during the low-hard state was
stable  characterized by a hard spectrum (with the power-law index
of $\sim$2, and the total flux in the power-law component being
$\sim$80\%) and 0.5 -- 10 Hz QPOs
(Trudolyubov et al. 1999; Muno et al. 1999). 
The fact that the canonical 
low-hard states
of black hole candidate sources have a negligible thermal component
(Chitnis et al. 1998) prompted Trudolyubov  et al. (1999) to characterize this
state as an ``intermediate state'' and they conclude that with the
lowering of the accretion rate the source should go to the canonical
hard state. Since the source was in similar state on several 
occasions (1996 July-August; 1997 October; 1998 September-October),
we treat this state as the ``low-hard state'' of GRS-1915+105.

After 1997 April 25 the source started a steady increase in its
X-ray emission with an average increase in the ASM count rate
of 0.65 s$^{-1}$ day$^{-1}$ and reaching a count rate of 76 s$^{-1}$
in the middle of July (day number 1290). The variability, as
can be deduced from the ASM count rates and measured as the
fraction of rms to mean, steadily increased
from 15\% to 40\%. It should be noted that during the
low-hard state of 1996 July-August the source showed similar 
variability behavior: the rms variation was low (5 - 10\%)
during the low-hard state and it was $\sim$40\% just before
and after this state (Paul et al. 1998b). 
The variability, however, decreased to $<$15\% around 1997 June
(Day number 1260). During this period the source showed evidences
of continuous ``ringing'' flares (Vilhu and Nevalainen 1998; Yadav et al. 1999),
with time scales of 30 -- 60 s and these will not be evident in 
the ASM data.
The variability increased again from June end (Day number 1275)
and the flaring state continued for about 20 more days with the
ASM variability being 30 - 40\%. The source reached a steady state 
with low variability ($\sim$10\%). The ringing flares started again in
the beginning of 1997 August (Yadav et al. 1999) and towards the end of
this state the peculiar outbursts accompanied by infrared flares 
were observed on 1997 August 14-15 (Eikenberry et al. 1998). 
The source was relatively quiet thereafter for about 10 days and 
then went into a flaring state. On September 9 the peculiar 
``outburst'' was seen again (Markwardt et al. 1999).

For a given accreting system, it is reasonable to assume that the 
long term changes in the X-ray emission is due to changes in the
accretion rate. Unlike Cygnus X-1 where the source transition
typically takes place in a few days, the state transition in
GRS~1915+105 took about three months. If we attribute the state 
transition as due to a change in the accretion rate near some
critical accretion rate (as has been done in Chakrabarti \& Titarchuk
1995), the slow state transition can be used to study the system in
detail at this critical rate. In this scenario the variety of 
bursts and variability seen in GRS~1915+105 
can be attributed to phenomena near this critical
state. Here we address the question of the fast state transition
that has occurred in 1997 June 18. From Figure 1 we can see that the
bursting properties systematically changed during the transition. In
Yadav et al. (1999) (see their Figure 2) it was shown that for about a
month the source was in a burst mode with a slow transition from
regular bursts to irregular bursts and then again to a 
regular burst of shorter duration.  
 In Yadav et al. (1999) it was postulated that the irregular bursts are
manifestations of rapid state changes.  To quantify the properties
of the spectral states we have taken two RXTE observations each
in the low-hard and high-soft states. In the
low-hard state the observation carried out on 1997 March 26 (towards the 
end of the low-hard state)
and  another on 1997 May 8 (in the beginning of state transition) were
selected. In the high-soft state observations carried out on 1997 July 20
and 1997 August 19 were chosen. The 1997 May 8 and 1997 August 19 
observations are the same that are used in Yadav et al. (1999). 
The X-ray emission properties of the source in the spectral states
as deduced from these observations are compared with that obtained
during the irregular bursts observed on 1997 June 18. Some salient
features of the selected observations are given in Table 1.
We use the standard 2 mode data which gives 128 channel energy
spectra every 16 s (used for the spectral analysis) and 
2 channel binned data below 13 keV every 0.125 ms (used for the 
timing analysis).

\subsection{The low-hard and the high-soft states of GRS~1915+105}

The light curves of the selected low-hard and high-soft states of the
source are shown in Figure 2, with a time resolution of 1 s. The
data is for all the five Proportional Counter Units (PCUs) and the
energy range is 2 $-$ 13 keV. The average count rate during the low-hard
state of March 26 is 3484 s$^{-1}$ and it increased to 5600 s$^{-1}$
for the May 8 data. During the high-soft states of July 20 and August 19,
the observed count rates were 18330 s$^{-1}$ and 21669 s$^{-1}$,
respectively. 

\begin{figure}
\centering
\psfig{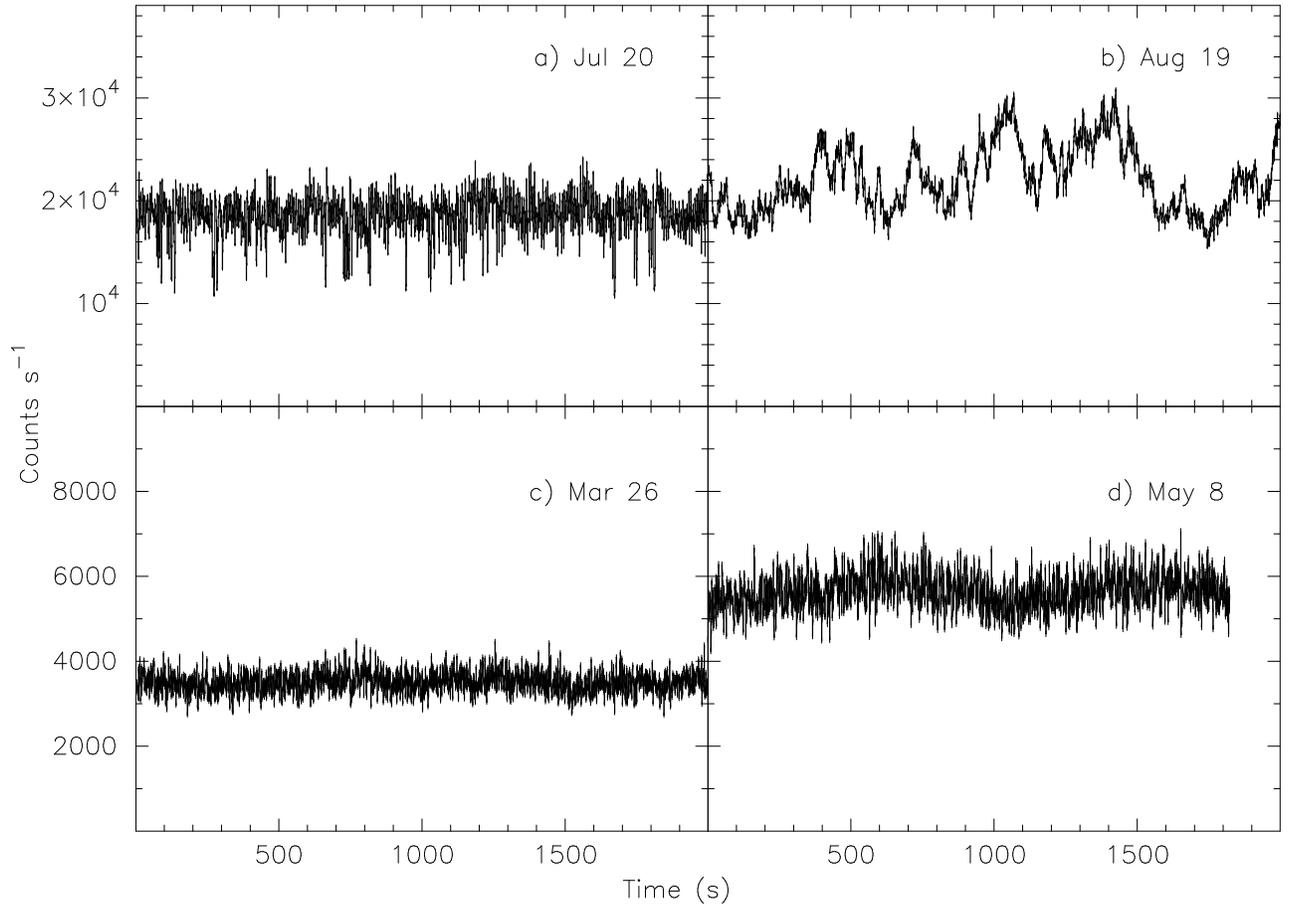}
\caption[fig2.eps]{
The 2 $-$ 13 keV X-ray light curve obtained form the RXTE-PCA 
is shown with a bin size of 1 s, for the four selected observations.
The observations during the high-soft state are shown in the
top two panels and those during the low-hard states are shown in
the bottom two panels. }\label{fig2}

\end{figure}

\placefigure{fig2}

 Several features of the state changes in black hole candidates are discernible
from the light curve itself. On March 26 (when the count rate was low) there
was higher variability in the 1 $-$ 10 s time scale but the source
was stable at longer time scales. The situation was opposite in the
high state of August 19: there is large variability at longer time scale.
On May 8 the count rate increased by about 50\%, but the variability 
characteristics were essentially the same. On July 20, though the 
count rate was similar to that of August 19, the light curve appears similar
to that seen during the low state, with several additional spikes at
a few tens of seconds. The behavior of the source is similar to the
Galactic black hole candidate source Cygnus X-1 (see Rao et al 1998) 
when the source showed essentially similar pattern of changes.

\begin{figure}
\centering
\psfig{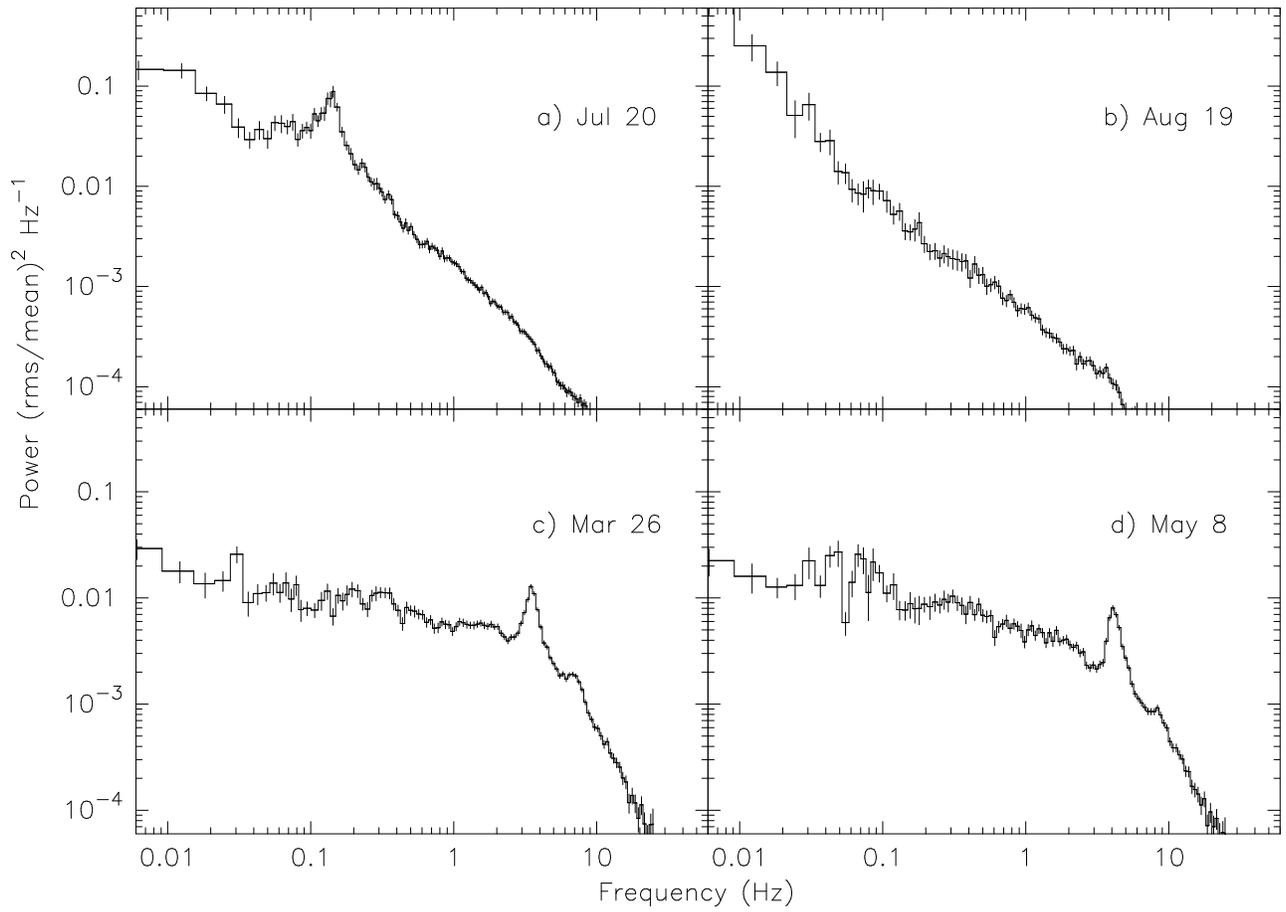}
\caption[fig3.eps]{
The power density spectrum (PDS) of GRS~1915+105 for 
the four observations shown in Figure 2.  }\label{fig3}
\end{figure}

\placefigure{fig3}

\begin{deluxetable}{lcccccc}
\footnotesize
\tablecaption{Details of the Power Density Spectrum (PDS) for
GRS~1915+105 \label{tbl-2}}
\tablewidth{0pt}
\tablehead{
\colhead{Obs. Date} &  \colhead{} &   \colhead{0.1 $-$ 1 Hz } & \colhead{} & \colhead{}  & \colhead{1 $-$ 10 Hz} & \colhead{}  \\
\colhead{} &  \colhead{Power-law} &   \colhead{Variability} & \colhead{QPO frequency} &   \colhead{Power-law} &   \colhead{Variability} & \colhead{QPO frequency}  \\
\colhead{} &  \colhead{Index} &   \colhead{(\% rms)} & \colhead{(Hz)} &   \colhead{Index} &   \colhead{(\% rms)} & \colhead{(Hz)}  }
\startdata
{\bf Low-hard  state} & & & & & & \nl
& & & & & & \nl
$~~~~~~$ Mar 26 & -0.32$\pm$0.05 & 8.0 & $-$  & -1.7$\pm$0.2  &  17.0  &  3.5  \nl
$~~~~~~$ May 8  & -0.37$\pm$0.07 & 7.7 & $-$  & -1.08$\pm$0.07  &  14.5  &  4.2  \nl
$~~~~~~$ Jun 18 (low) & -0.5$\pm$0.2 & 5.1 & $-$  & $<$-1.8  &  11.0  &  3.3  \nl
& & & & & & \nl
{\bf High-soft state} & & & & & & \nl
& & & & & & \nl
$~~~~~~$ Jun 18 (high)$^a$ & -1.3$\pm$0.1 & 3.3 & $-$  & -1.3$\pm$0.1  &  2.6  &  $-$ \nl 
$~~~~~~$ Jul 20 & -1.3$\pm$0.1 & 9.0 & 0.13 & -1.61$\pm$0.02  &  4.8  &  $-$ \nl
$~~~~~~$ Aug 19$^a$  & -1.25$\pm$0.02 & 3.6 & $-$  & -1.25$\pm$0.02  &  3.0  &  $-$ \nl
\enddata
\tablenotetext{a}{Power-law is fitted for the complete frequency range of 0.1 -- 10 Hz}
\end{deluxetable}

 To quantify the timing characteristics, we have obtained the power density
spectrum (PDS) for these four observations and these are displayed in
Figure 3. 
The low-hard state PDS displays all the characteristics of PDS generally
observed for other Galactic black hole sources during the low-hard state
(see Rao et al. 1998 and references therein): a flat spectrum up to a
break frequency in the vicinity of a few Hz
and spectrum falling steeply thereafter. 
The  narrow quasi-periodic
oscillation (QPO) along with its first harmonic, seen during the other 
low-hard states of GRS 1915+105 (Paul et al. 1997; Morgan et al. 1997),
is evident in the PDS. 
In the high state the PDS is a featureless
power-law above 0.2 Hz.  In the high state of July 20, however, there
is a low frequency QPO at 0.13 Hz. The salient features of the
PDS in the two states are given in Table 2. To highlight the
differences in the PDS at the low and high frequencies, we have fitted
the PDS with a power-law and a Lorentzian (whenever a QPO is present) in
the frequency ranges 0.1 $-$ 1 Hz and 1 $-$ 10 Hz. 
For the high state data a single power-law was fitted in the entire
frequency range.
The distinguishing
features of the spectral states are 1) a flat spectrum below 1 Hz for the
low-hard state with the
power-law index ranging from -0.5 to -0.2 compared to an index 
$\sim$ -1.3 in the high-soft state, 2) presence of a narrow 0.5 --  10 Hz 
QPO feature in the low-hard state and 3) higher variability ($\sim$15\%)
in the 1 $-$ 10 Hz range in the low-hard state compared to the low
variability ($<$5\%) in the same frequency range for the high-soft state.

The other distinguishing characteristics of the intensity states
of Galactic black hole candidate sources is  the energy spectra.
The low state is dominated by a thermal-Compton spectrum
(which can be approximated to a power-law at lower energies)
along with a blackbody emission component, which generally is modeled
as a disk blackbody emission. The disk blackbody emission increases
in intensity and dominates the spectrum in the high state.

\placetable{tbl-2}

We have generated the 128 channel energy spectra from the Standard 2 mode
of the PCA for each of the above observations. Standard procedures for
data selection, background estimation and response matrix generation
have been applied. To avoid the extra systematic errors in the 
response matrix of PCA, we have restricted our analysis to the
energy range of 3 $-$ 26 keV. Data from all the PCUs are added together.
We have fitted the energy spectrum of the source  using a model consisting
of  disk-blackbody and  power-law with absorption by intervening cold 
material parameterized as equivalent Hydrogen column density, N$_H$.
The value of N$_H$ has been kept fixed at 6 $\times$ 10$^{22}$ cm$^{-2}$.
We have included a Gaussian line near the expected K$_\alpha$  emission
from iron and absorption edge due to iron. These features help to mimic
the reflection spectrum usually found in other Galactic black hole
candidate sources like Cygnus X-1 (Gierlinski et al. 1997).
Systematic errors of 1\% have been added to the data. We obtain
reduced $\chi^2$ values in the range of 1 -- 3.

\begin{figure}
\centering
\psfig{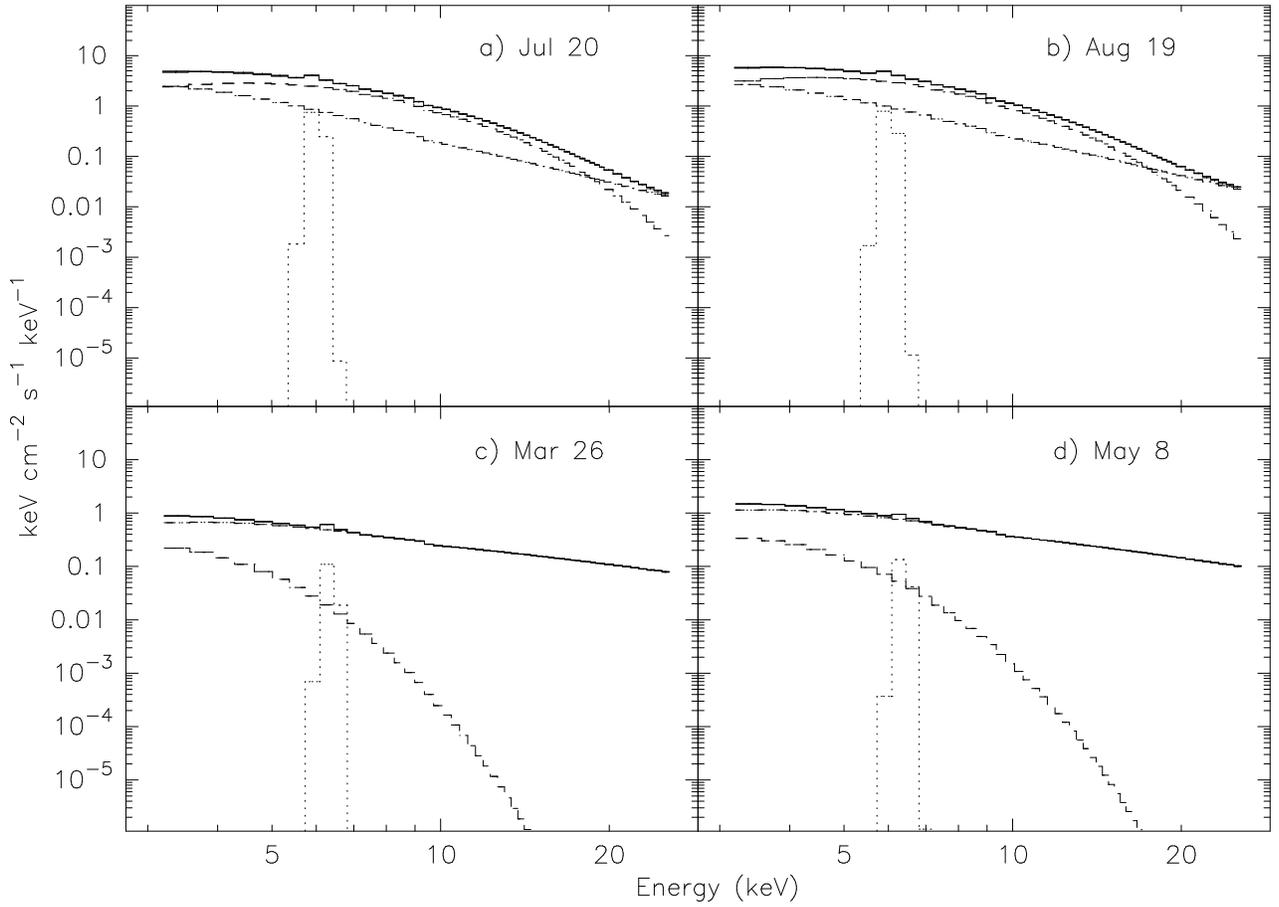}
\caption[fig4.eps]{
The 3 $-$ 26 keV unfolded energy spectra of GRS~1915+105 obtained
from the four selected observations.  The spectra are fitted with a model
consisting of a disk blackbody, powerlaw along with absorption
due to neutral matter. An iron line and iron edge are also included in the
model. }\label{fig4}
\end{figure}

\placefigure{fig4}

The unfolded spectrum
for the two spectral states are shown in Figure 4.  The individual model 
components are shown separately. Scales for the axes are kept the same
for all the plots. The derived parameters
are given in Table 3.  The quoted errors are nominal 90\% confidence levels
obtained by the condition of $\chi_{min}^2$ + 2.7. The inner disk radius
is calculated using an inclination angle of 70$^\circ$ for the accretion
disk and a distance of 10 kpc to the source (see Muno et al. 1999).
The iron line was found to
have an equivalent width of $\sim$100 eV.

\placetable{tbl-3}

\begin{deluxetable}{llllcc}
\footnotesize
\tablecaption{Details of the spectral fits \label{tbl-3}}
\tablewidth{0pt}
\tablehead{
\colhead{Obs. Date}  &   \colhead{kT$_{in}$} & \colhead{R$_{in}$} &   \colhead{ $\Gamma$} &   \colhead{F$_{bb}$ } &   \colhead{F$_{PL}$} \\
\colhead{} & \colhead{(keV)}  &   \colhead{(km)} &   \colhead{} &   \colhead{} &   \colhead{} }
\startdata
{\bf Low-hard  state} &  & & & & \nl
& & & & & \nl
$~~~~~~$ Mar 26 & 0.71$\pm$0.07 & 68 & 2.37$\pm$0.02  & 0.06  &  0.88  \nl
$~~~~~~$ May 8  &  0.84$\pm$0.08 & 52 & 2.52$\pm$0.02  & 0.11  &  1.34  \nl
$~~~~~~$ Jun 18 (low) &  0.92$\pm$0.04 & 55 & 2.38$\pm$0.05  & 0.23  &  0.59  \nl
& & &  & & \nl
{\bf High-soft state}  & & & & & \nl
& & & &  & \nl
$~~~~~~$ Jun 18 (high)& 2.14$\pm$0.03 & 21 & 2.93$\pm$0.05  & 3.45  &  1.59  \nl
$~~~~~~$ Jul 20 & 2.05$\pm$0.03 & 21 & 3.81$\pm$0.05  & 2.58  &  1.27  \nl
$~~~~~~$ Aug 19 &  1.96$\pm$0.02 & 26 & 3.69$\pm$0.05  & 3.21  &  1.46  \nl
\enddata
\tablenotetext{}{{\bf Note:} The model consists of (see text) a disk-blackbody (with inner disk 
temperature kT$_{in}$ and radius R$_{in}$ as parameters) and a powerlaw (with photon index $\Gamma$). 
The 3 $-$ 26 keV blackbody and powerlaw observed fluxes 
(F$_{bb}$ and     F$_{PL}$, respectively) are given in units of 
10$^{-8}$ ergs cm$^{-2}$ s$^{-1}$. } 
\end{deluxetable}

We can identify the following distinguishing features in the
energy spectrum in the two spectral states of the source: 1) in
the low-hard state the disk blackbody component has lower temperature
($<$1 keV) and larger inner disk radius ($>$50 km) compared to the
high-soft state, which has the inner disk temperature of $\sim$2 keV and
inner disk radius of $\sim$20 km. 2) the disk blackbody component has a 
3 $-$ 26 keV flux of $\sim$10$^{-9}$ erg cm$^{-2}$ s$^{-1}$ ($<$10\% of
the total flux) whereas the disk blackbody flux increases by a factor of
more than 30 in th high-soft state. In fact it becomes the predominant component
with $>$65\% of the observed flux being in this component. 3) The
power-law index becomes noticeably steeper in the high-soft state.

\subsection{Rapid state transitions }

Now we show that during the irregular bursts observed on 1997 June 18
the source GRS~1915+105 made rapid intensity transitions between two 
levels and the X-ray emission properties in these two levels are
identical to the spectral states of the source.
Belloni et al. (1997b) have shown that during the irregular bursts 
observed on 1997 June 18 the source
exhibited a repeating pattern of intensity states characterized by
well defined spectral states. 
We have selected one   irregular burst from this observation
for the timing and spectral analysis. 
 The power density spectrum (PDS) were separately
obtained for the burst and quiescent time. 
The results are shown in Figure 5, right panel, separately for quiescent data 
and the burst data. The results are also given in Table 2.
The results of the spectral analysis is shown in the left panels of the
same figure and the derived parameters are given in Table 3.
 The scales for figures for the PDS and the energy spectra 
 are identical to Figure 3 and Figure 4, respectively.

\begin{figure}
\centering
\psfig{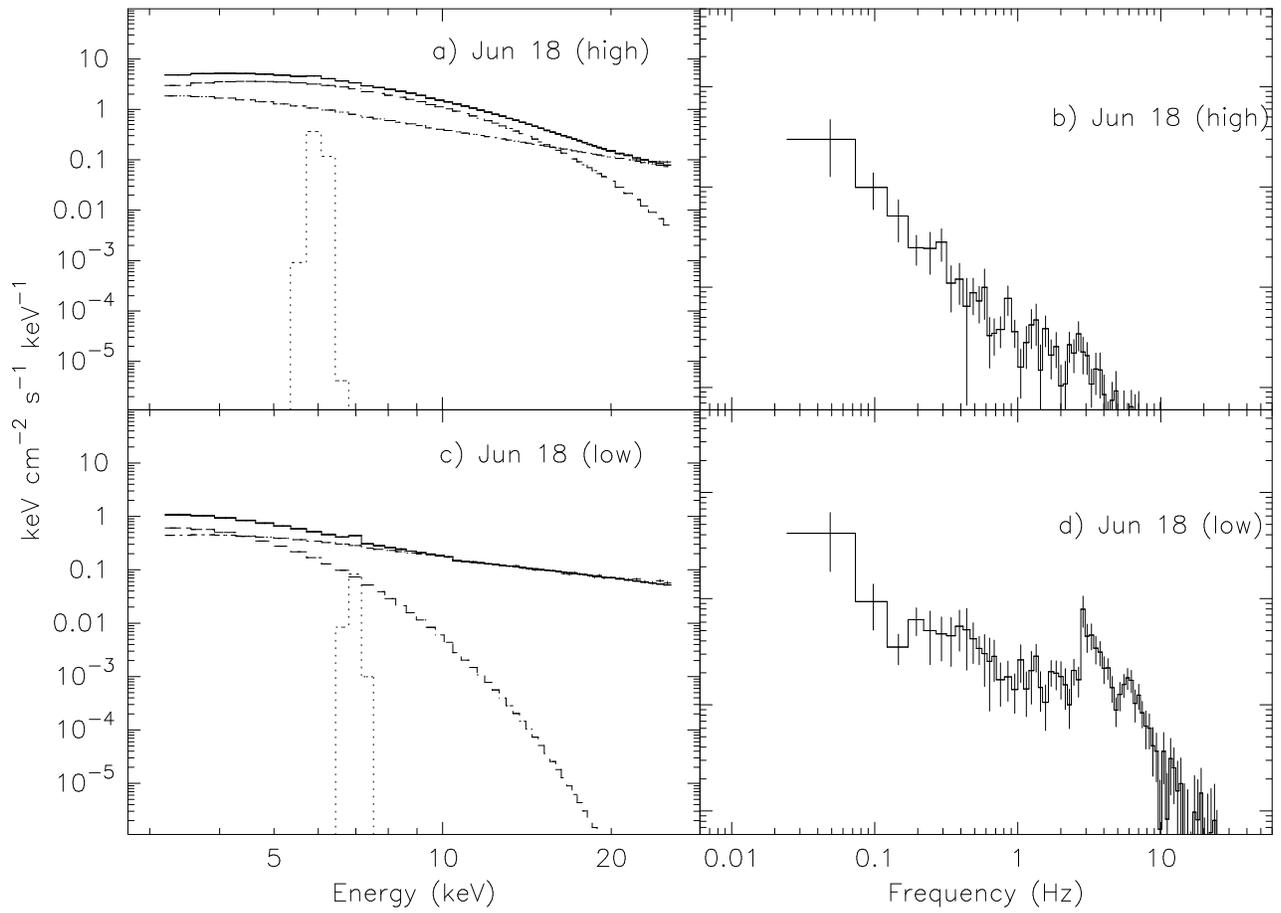}
\caption[fig5.eps]{
The power density spectrum (PDS) and the energy spectrum of GRS~1915+105 
obtained during the state transition observed on 1997 June 18. The scales
for figures for the PDS and the energy spectra 
 are identical to Figure 3 and Figure 4, respectively.  }\label{fig5}
\end{figure}

\placefigure{fig5}

 The remarkable similarity between the low-hard state and burst quiescence
is evident from the figure. They show a power-law
spectral index (between 0.1 and 1 Hz) of -0.3 and -0.5
respectively. A QPO feature at $\sim$4 Hz is seen during the
low-hard state and there is an indication of a narrow QPO feature at
3.2 Hz during the burst quiescence.
 The observed PDS characteristics are very similar during the high-soft
state of 1997 August 19 and the  burst time observations of 1997 June 18.
The energy spectra, too, highlights the similarity between the spectral state
changes and the bursts (see Table 3 and Figure 5). All the identifying 
features of the spectral states like the changes in the
disk blackbody temperature, radius and flux are evident in the
burst time too.

 These observations strongly support the suggestion, made in
Yadav et al. (1999), that the source makes rapid transition between
the low-hard state and high-soft state in very short time scales.
The difference in the PDS is predominantly above 1 Hz and hence
to dramatize our findings we have plotted the observed sub-second
variability in these two states.
  The sub-second variability is
characterized as the rms (root mean square) normalized to mean
for 10 bins of 0.1 s duration. The contribution from counting statistics
has been subtracted. In Figure 6, the observed count rates
and the variability are plotted. The irregular
burst described earlier is shown in Figure 6a (upper-left panel)
and the corresponding variability is shown in Figure 6b (lower-left
panel). The rms variability makes a transition when the source
count rate changes. The observed count rates during the 
low-hard state (1997 May 8 observations) and the high-soft state (1997
August 19 observations, shown as stars) are shown in Figure 6c. 
The variability values for these two states are plotted in Figure 6d,
with the high-soft state shown as stars. The similarity 
between the burst quiescence and the low-hard state on the
one hand and the burst and the high-soft state is very striking.
The total count rate (in 2 $-$ 13 keV range) is 3000 s$^{-1}$ and the
variability is 9.3\% during the burst quiescence, which 
compares well with the values of 5400 s$^{-1}$ and 12.9\%, 
respectively, observed during the low-hard state of 1997 May.
The source had a count rate of 19000 s$^{-1}$ and variability
of 3.4\% during the high-soft state of 1997 August, which is
distinctly different from the values found during the low-hard
state but very similar to the values of 25000 s$^{-1}$ and 
2.9\%, respectively, found during the burst.
 
\begin{figure}
\centering
\psfig{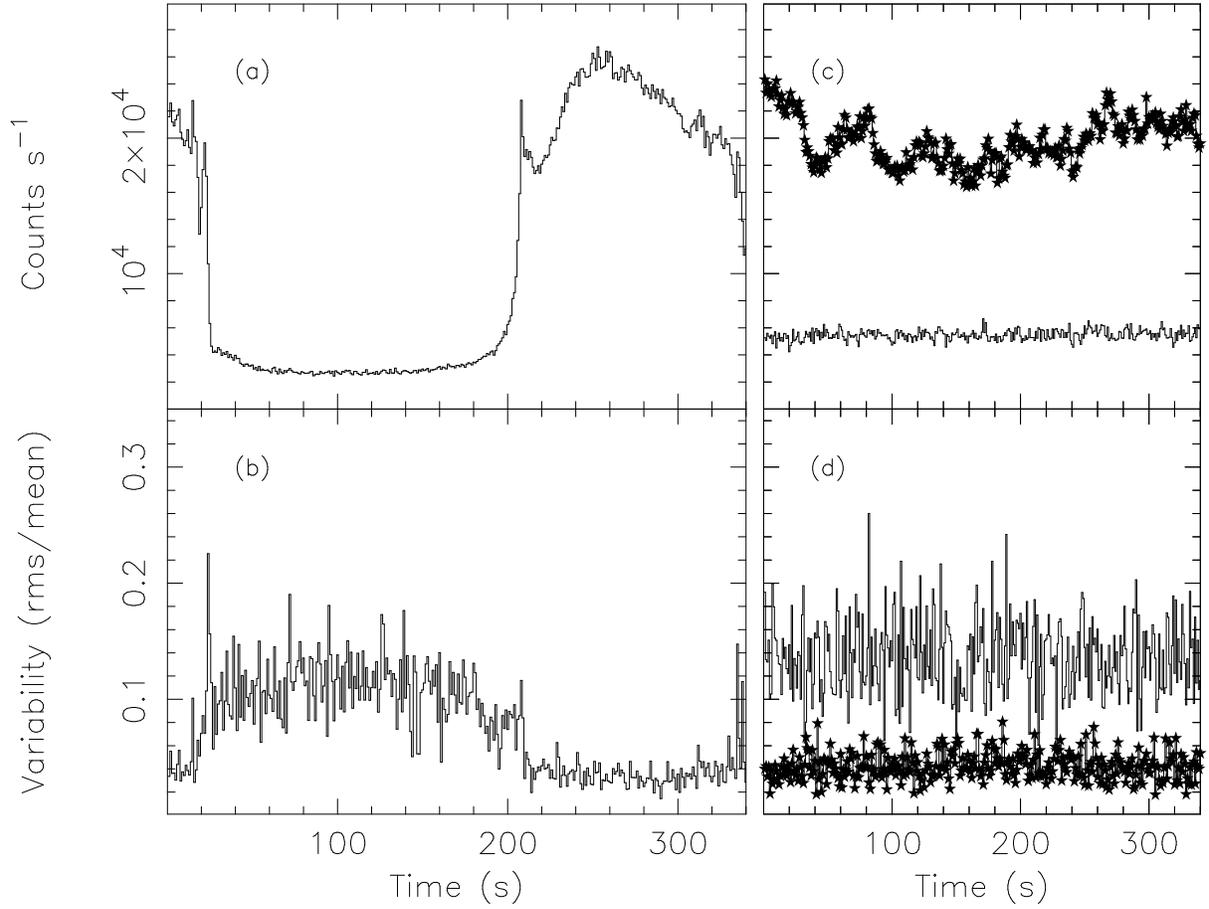}
\caption[fig6.eps]{The sub-second variability (defined as the rms for 10 bins
of 0.1 s duration normalized to the mean) is shown for GRS 1915+105 along with
the observed count rates obtained from the RXTE PCA data. The contribution
from counting statistics has been subtracted. One irregular burst observed on
1997 June 18 is shown in Figure 6a and the corresponding variability is shown
in Figure 6b. The observed count rates during the low-hard state (1997 May 8
observations) and the high-soft state (1997 August 19 observations, shown as
stars) are shown in Figure 6c. The variability values for these two states
are plotted in Figure 6d, with the high-soft state shown as stars.}
\label{fig6}
\end{figure}

\placefigure{fig6}

To highlight the spectral similarities we have shown the spectral 
ratios in Figure 7.
Four spectral files are generated. 
Quiescent data for the irregular burst were accumulated in a
spectral file, referred
to as B$_Q$, for burst quiescence. Similarly burst data for the 
same burst  is accumulated in the file  B$_H$ (burst-high).
Spectral data obtained on 1997 May 8 (during the low-hard state
of the source) are accumulated in the file L$_H$ 
and the high-soft state data (1997 August 19)
 was accumulated in the file H$_S$. 

\begin{figure}
\centering
\psfig{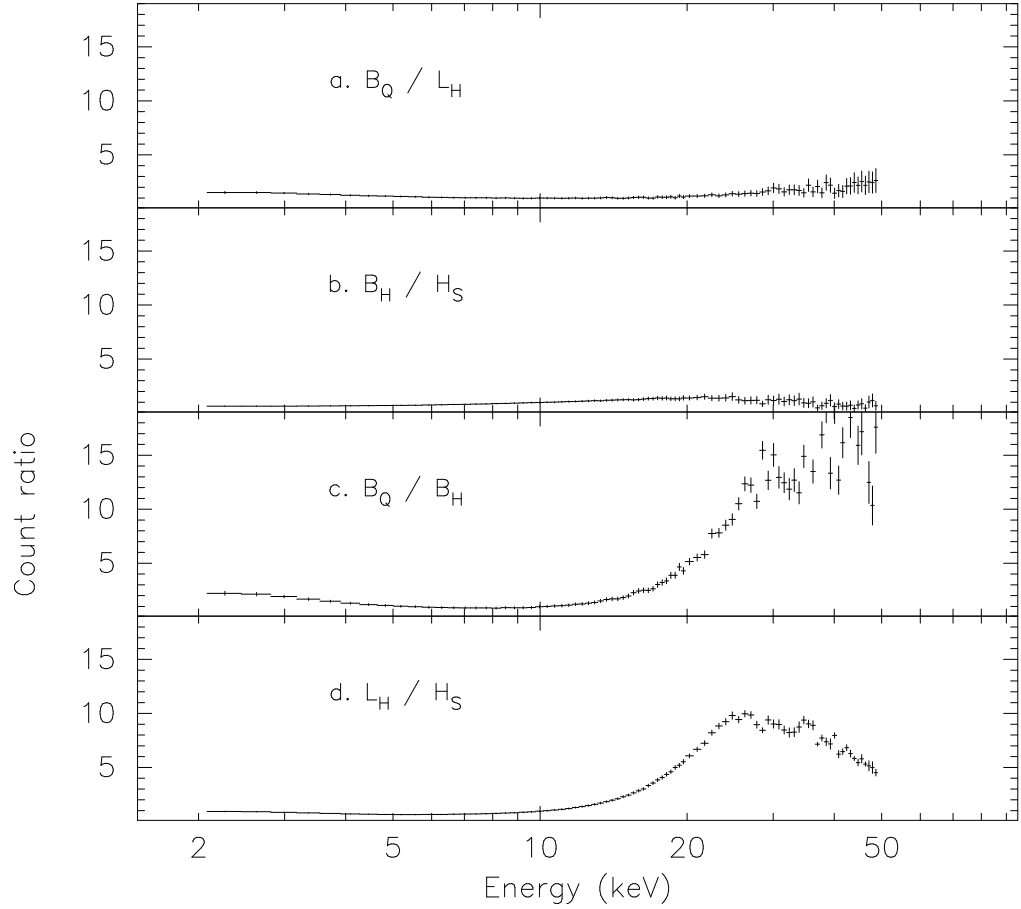}
\caption[fig7.eps]{
Spectral ratio plots for the different spectral states  are shown here.
The ratio of Burst quiescence to low-hard state spectrum
(B$_Q$ to L$_H$ ratio), as a function of energy, is plotted in
Figure 7a, and the burst to high-soft state spectral ratio 
(B$_H$ to H$_S$) is shown in Figure 7b. All the spectra
are normalized to the observed values at 10 keV. The spectral ratio seen
during the burst (B$_Q$ to B$_H$) is plotted in Figure 7c and,
the ratio of the Low-hard to  High-soft spectra (L$_H$ to H$_S$)
  is given in Figure 7d.}
\label{fig7}
\end{figure}

\placefigure{fig7}

The ratio of B$_Q$ to L$_H$, as
a function of energy, is plotted in Figure 7a, 
and B$_H$ to H$_S$ ratio is shown in Figure 7b. All the
spectra are normalized to the observed values at 10 keV.
It can be seen that the spectra during the burst  quiescence
is very similar to that during the
low-hard state of the source and burst spectrum is
very similar to that seen during the high-soft state of the
source, within a factor of two. To emphasize the spectral
change during the burst the spectral ratio seen during the burst 
(B$_Q$ to B$_H$) is plotted in Figure 7c and, for
comparison, the ratio of the low-hard  to high-soft spectra 
are plotted in Figure 7d. It can be seen that the burst quiescence
spectrum is very different from the burst spectrum
characterized by a hard component above 10 keV. This type
of a hard component is evident in the low-hard state spectrum
as compared to the high-soft state (Figure 7d). Hence we
conclude that during an irregular burst the source shows
a spectral change which is quite similar to the spectral change
seen during the transition from a low-hard state to a 
high-soft state.

\section{DISCUSSION}

 Though GRS 1915+105 defies a neat classification between low-hard
and high-soft state as seen in other Galactic black hole candidate
sources, it does exhibit extended low-hard states (\cite{grei:98};
\cite{morg:97}). Outside such extended low-hard states the
source can be further sub-divided into several states namely 
bright state (high-soft state),  chaotic state and flaring state
(\cite{morg:97}). We have shown here that in at least one type 
of flaring state when the source was showing irregular bursts, the
source traverses between low-hard state and the high-soft state.

The PDS presented here during the burst quiescence and the low-hard
state are similar to that seen during the extended low-hard state of
1996 July-August (\cite{morg:97}; \cite{paul:97}). The PDS is flat up to
a few Hz with evidence of QPO in the range of 0.5 to 5 Hz. The 
average power in the range of 0.1 to 1 Hz presented here is 0.01 (rms/mean)$^2$
Hz$^{-1}$ which is consistent with the earlier observations. The PDS during
the burst and the high-soft state are similar to that seen during the
bright state of 1996 April (\cite{morg:97}) with the PDS being very steep
between 0.1 and 5 Hz.

 We emphasize here that the high-soft and low-hard states of the Galactic black hole candidate
sources are generally identified only
from the strength and shape of the spectra. Since we see
spectral changes in a very short time we can conclude that
the canonical states too  change in such time scales. 
The timing characteristics like the shape of PDS, QPO, and rms
variations give additional support to this.
This reinforces the conclusion drawn earlier 
(Yadav et al. 1999) that the source can make very rapid state transitions. 

There have been attempts to explain the rapid variability seen in 
GRS 1915+105 using disk instability models. Taam et al. (1997)
have used scaling laws for viscosity and found that the instability
time scales of the inner accretion disk can explain the time scales
of the regular bursts seen in GRS 1915+105. Belloni et al. (1997b)
have tried to explain the repeated patterns as due to the appearance
and disappearance of inner accretion disk. 
Nayakshin, Rappaport, \& Melia (2000)   have investigated the different accretion models and
viscosity prescriptions and attempted to explain the temporal behavior of
GRS 1915+105. In particular, they have shown that the accretion instability 
in a slim disk, as invoked by Taam et al. (1997), is not likely to 
adequately account for the behavior of GRS 1915+105 (because of the difficulty
in maintaining the high state for long) and the appearance of inner 
accretion disk, as postulated by Belloni et al. (1997), will
take a time scale comparable to the burst time scale. Though Nayakshin
et al. (2000) were able to reproduce many characteristics in the
X-ray variability for GRS~1915+105, they were unable to explain the
rise/fall times or the $<$ 10 s oscillations. 

 The results presented in this paper, namely, the occurrence of fast
changes in the luminosity state, put additional constraints to any disk
instability model. The fact that the spectral states are observed
for long durations 
show that these states are stable solutions for an
accretion disk rather than changes in any stability parameters. 
Such stable solutions changing in a short time scale
is difficult to explain in terms of the
disappearance of slim disks. Hence in Yadav et al. (1999) it was postulated that
these changes are due to appearance and disappearance of sub-Keplerian
disks above the slim disks. 

Recently, Chakrabarti (1999) have presented a possible solution to the
state transition based on a Two Component Accretion Flow (TCAF)
model given
by Chakrabarti \& Titarchuk (1995). They have invoked the initiation of
mass outflow (see also Das \& Chakrabarti 1999) 
as the cause of the start of the state change and the catastrophic
Compton cooling of the material in a spherical volume causing the
reverting back to the high state. This model essentially reinforces the
suggestion of Yadav et al. (1999) but gives a physical basis for the
start of the event. In a later work, Chakrabarti \& Manickam (2000)
have expanded this model further and derived a correlation between
the burst time and the QPO frequency. It would be interesting
to make a detailed time resolved X-ray spectroscopy during a regular
burst to find the slow evolution of the optical depth of the Compton 
cloud.

Since the canonical spectral states of black hole binaries
remain for considerably long duration (up to years)
in the same state with similar properties, it is usually
assumed that they are some stable solutions for the accretion disk.
It is normally believed that for a given source the total
accretion rate is the governing parameter for the spectral changes.
Our finding that
the source makes very fast repeated state transitions, implies
that the two solutions for the accretion disk (corresponding to
the two spectral states) must exist for roughly the same total
mass accretion rate because the time scale of the state
change ($\sim$10s) observed during the irregular bursts
is much smaller than the time scale for the readjustment of the
accretion disk for any global mass accretion rate changes. The  time
scale for the latter must be of the order of thousand seconds because it is normally
believed that far away from the compact object the accretion disk
can be described as the classical thin disk which has a
very large  viscous time scale. Hence two solutions for
the accretion disk must exist for the same global accretion rate.

Such behavior of fast spectral changes must be occurring in other black
hole binaries where a wide range of X-ray luminosities are observed,
like GRO J1655-40. 
Remillard et al. (1999) have pointed out the similarity between
the temporal properties of GRS 1915+105 and GRO J1655-40
when each source  teeters between relative stability and a 
state of intense oscillations. 
It will be very interesting to find
long duration irregular bursts in GRO J1655-40 because in this source
the binary period and
the mass of the black hole are known and hence it will
prove an ideal ground to further refine the TCAF model.

\section{CONCLUSION}

We have presented the X-ray spectral and time variability characteristics
of GRS 1915+105 when the source was exhibiting irregular bursts. The spectral
characteristics strongly suggest the view that the source makes a state
transition in a very fast time scale (a few seconds). We have found that the
temporal variabilities also support this conclusion.  We
conclude that the source can make spectral state transition only if the
advective and standard thin disk co-exist, as suggested by \cite{chak:95}.

\begin{acknowledgements}
\acknowledgments

This research has made use of data obtained through the High Energy
Astrophysics Science Archive Research Center Online Service, provided by the
NASA/Goddard Space Flight Center.

\end{acknowledgements}

\clearpage

\end{document}